\documentclass[aps,prd,twocolumn,showpacs,nofootinbib]{revtex4-1}

\usepackage{graphicx,amsfonts,amssymb,amsbsy}
\usepackage{pstricks}
\usepackage{appendix}
\usepackage{graphicx}
\usepackage{slashed}
\usepackage{bm}
\usepackage{amssymb}
\usepackage{amsmath}
\usepackage{tabularx}
\usepackage{natbib}

\begin{document}

\title{Constraining color flavor locked strange stars in the gravitational wave era} 

\author{C. V\'asquez Flores and G. Lugones}

\affiliation{Centro de Ci\^{e}ncias Naturais e Humanas, Universidade Federal do ABC, \\ Av. dos Estados 5001, CEP 09210-580, Santo Andr\'{e}, SP, Brazil}

\begin{abstract}
We perform a detailed analysis of the fundamental mode of non-radial pulsations of color flavor locked strange stars. Solving the general relativistic equations for non-radial pulsations for an equation of state derived within the MIT bag model, we  calculate the frequency and the gravitational damping time of the fundamental mode for all the parametrizations of the equation of state that lead to self-bound matter. 
Our results show that color flavor locked strange stars can emit gravitational radiation in the optimal range for present gravitational wave detectors and that it is possible to constrain the equation of state's parameters if the fundamental oscillation mode is observed and the stellar mass is determined.
We  also show that the $f$-mode frequency can be fitted as a function of the square root of the average stellar density $\sqrt{M/R^3}$ by a single linear relation that fits quite accurately the results for all parametrizations of the equation of state.   All results for the damping time  can also be fitted as a function of the compactness $M/R$  by a single  empirical relation. Therefore, if a given compact object is identified as a color flavor locked strange star these two relations could be used to determine  the mass and the radius from the knowledge of the frequency and the damping time of gravitational waves from the $f$ mode.  
\end{abstract}

\pacs{97.60.Jd, 26.60.Kp, 97.10.Sj, 95.85.Sz, 04.40.Dg}

\maketitle

\section{Introduction}

Very recently, gravitational waves were detected at the Advanced LIGO interferometer \cite{LIGO}, opening a new window for the observation of the Universe.  When the design sensitivity is reached by the second generation of interferometric gravitational wave detectors (Advanced LIGO, Advanced Virgo and KAGRA),  which is expected to happen around the year 2021, such instruments will be ten times more sensitive than the first generation. Such sensitivity will probably allow the detection of pulsation modes of compact stars excited in binary mergers,  in newly born  compact objects associated with the violent dynamics of core collapse supernovae, or due to the structure rearrangements induced by phase transitions at neutron star cores.

The vast family of oscillation modes of compact stars can be classified into  polar  modes  producing spheroidal fluid deformations, and  axial modes producing toroidal deformations. Modes are said to be fluid if they involve a significant fluid motion, and spacetime if they  only involve vibrations of spacetime. In the case of polar fluid modes, the most well know in the literature are  the  fundamental ($f$), the pressure ($p$), and  the gravity ($g$) modes, but there are other relevant modes such as the shear mode in a solid crust or  the Alfv\'en modes in the presence of magnetic fields. In the case of axial modes, there are  torsional modes  and rotational modes (Rossby and  inertial modes). There are also polar and axial spacetime modes, which only involve vibrations of spacetime ($w$ modes).

The analysis of pulsation modes can reveal details about the internal properties of neutron stars 
\cite{detweiler1985,andersson1998,miniutti2003,benhar2004}.
In particular, there has been great interest in the identification of features in the modes that may characterize unequivocally  the appearance of deconfined quark matter \cite{benvenuto1998,yip1999,sotani2003,sotani2004,vasquez2010,sotani2011,vasquez2014}. In the present work we focus on the effect of color superconductivity on polar fluid modes, since pairing interactions are expected to have a key role in quark matter at sufficiently large densities and low temperatures.  
At asymptotically  large densities, where the quark masses are negligibly small compared to the quark chemical potential, three-flavor quark matter is in the color-flavor locked (CFL) state \cite{alford1999}. In this state quarks form Cooper pairs of different color and flavor where all quarks have the same Fermi momentum and electrons cannot be present \cite{Rajagopal2001}. Color-flavor locking has a significant effect on many features of quark matter, such as transport properties. 
In the equations of state (EOS), it introduces corrections of order $(\Delta/\mu)^2$ which is around  a few percent for typical values of the color superconducting gap ($\Delta \sim 0-150$ MeV) and the baryon chemical potential ($\mu \sim 300-400$ MeV). However, the effect is proportionally very large in the low pressure regime that affects the absolute stability of quark matter. Thus,  self-bound  stars  made up of quark matter from the center up to the stellar surface (\textit{strange stars}) may exist for a wide range of parameters of the MIT Bag model EOS \cite{lugones2002}. Studies of the structure of these objects show that color superconductivity  affects significantly the mass-radius relationship of strange stars, allowing for very large maximum masses \cite{lugones2003,lugones2004}.

In previous works we explored the impact of color superconductivity on radial oscillations  of color flavor locked strange stars \cite{vasquez2010}.  We also used the Cowling approximation to analyze the non-radial fluid modes of hadronic, hybrid and pure self-bound strange quark stars, including stars with color supercondicting phases  \cite{vasquez2014}.   
In the present work we concentrate on  the fundamental oscillation mode because it is the most efficient emitter of gravitational waves in the case of non-rotating objects. Specifically,  we concentrate on the effect that color superconductivity could produce in the eventually detectable gravitational wave signal related to the $f$-mode for stellar configurations with a maximum mass above  $2 M_{\odot}$ \cite{demorest2010,antoniadis2013}.

The paper is organized as follows: in Sec. \ref{section:EOS} we present the EOS for CFL strange quark matter. In Sec. \ref{oscillation} we present the equations that govern the non radial oscillations of neutron stars and discuss the numerical procedure used to solve the equations.  In Sec. \ref{results} present our results spanning all possible parametrizations of the  EOS for CFL strange quark matter and give analytic fittings for the frequency as a function of the average stellar density and for the gravitational damping time as a function of the stellar compactness.  Finally, in Sec. \ref{conclusions} we give a brief summary and discuss the astrophysical implications of the results.

\section{Thermodynamics of the CFL phase} 
\label{section:EOS}

\begin{table}[tb]
\centering
\begin{tabular}{c | ccc}
\hline \hline
Model    &    B  [MeV fm$^{-3}$]    &  $\Delta$ [MeV]  &  $m_s$   [MeV]  \\
\hline  
CFL1       &   60                 &   50       &      0      \\ 
CFL2       &   60                 &   50       &        150  \\
CFL3       &   60                 &   100       &     0        \\ 
CFL4       &   60       &   100       &     150   \\ 
CFL5       &   60       &   150       &     0        \\
CFL6       &   60       &   150       &     150    \\
CFL7       &   80       &   100       &     0    \\
CFL8       &   80       &   100       &     150    \\
CFL9       &   80       &   150       &     0    \\
CFL10       &   80       &   150       &     150    \\
CFL11       &   100     &   50       &        0       \\ 
CFL12       &   100     &   100       &     0         \\ 
CFL13       &   100     &   100       &     150    \\ 
CFL14     &   100     &   150       &     0        \\ 
CFL15     &   100     &   150       &     150    \\ 
CFL16       &   120       &   100       &     0    \\
CFL17       &   120       &   150       &     0    \\
CFL18       &   120       &   150       &     150    \\
CFL19       &   140       &   150       &     0    \\
\hline \hline 
\end{tabular}
\caption{Set of parameters employed in the present paper.} \label{models}
\end{table}

\begin{figure}[tb]
\includegraphics[angle=270,scale=0.35]{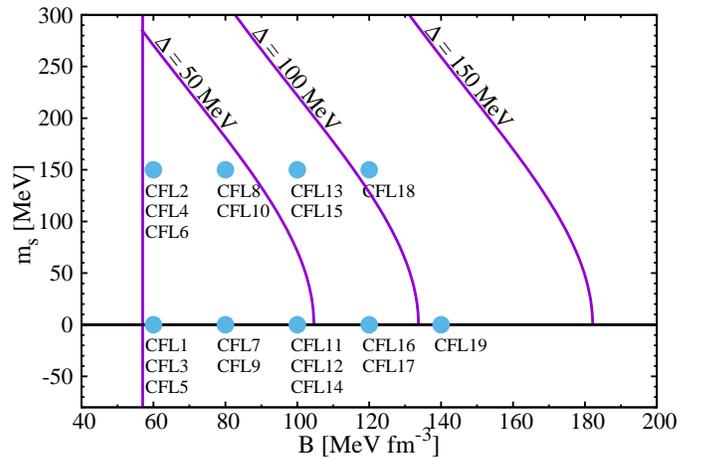}
\caption{Three stability windows of CFL quark matter, for $\Delta=50, 100, 150$ MeV. The vertical solid line is the limit imposed by requiring instability of two-flavor quark matter. For each of the indicated values of $\Delta$, CFL quark matter  is absolutely stable  if the strange quark mass $m_s$ and the bag constant $B$ lie inside the corresponding  bounded region. For example, a model with $B=100$ MeV fm$^{-3}$  and $m_s = 150$ MeV  is inside the stability window for $\Delta= 100, 150$ MeV  but is outside the stability window for $\Delta= 50$ MeV.}
\label{window}
\end{figure}

The equation of state for CFL quark matter can be obtained in the framework of the MIT bag model. To order $\Delta^{2}$ and  $m^{2}_{s}$ the pressure and energy density can be written as \cite{lugones2002}:
\begin{equation}\label{PB}
p=\frac{3\mu^{4}}{4\pi^{2}}+\frac{9\alpha \mu^{2}}{2\pi^{2}} - B,
\end{equation}
\begin{equation}\label{energiaB}
\epsilon = \frac{9\mu^{4}}{4\pi^{2}}+\frac{9\alpha\mu^{2}}{2\pi^{2}} + B,
\end{equation}
where
\begin{equation}\label{alfa}
\alpha = -\frac{m^{2}_{s}}{6}+\frac{2\Delta^{2}}{3}.
\end{equation}
From the above expressions we can obtain an analytic expression for $\epsilon = \epsilon(p)$: 
\begin{equation}\label{energiaC}
\epsilon = 3p+4B-\frac{9\alpha\mu^{2}}{\pi^{2}},
\end{equation}
with
\begin{equation}\label{mu2}
\mu^{2}= -3\alpha +\bigg[\frac{4}{3}\pi^{2}(B+p)+9\alpha^{2}\bigg]^{1/2}.
\end{equation}
In a similar way we can also obtain $p=p(\epsilon)$
\begin{equation}\label{PC}
p=\frac{\epsilon}{3}-\frac{4B}{3}+\frac{3 \alpha\ \mu^{2}}{\pi^{2}},
\end{equation}
with $\mu$ given by
\begin{equation}\label{mu2B}
\mu^{2}=-\alpha + \bigg[\alpha^{2}+\frac{4}{9}\pi^{2}(\epsilon-B)\bigg]^{1/2}.
\end{equation}

The adiabatic index $\gamma$ reads:
\begin{equation}\label{dp_de}
\gamma \equiv \frac{(\epsilon+p)}{p} \frac{dp}{d\epsilon}  = \frac{(\epsilon+p)}{p}   \left[  \frac{1}{3}+\frac{2\alpha}{3}\bigg(\frac{1}{\mu^{2}+\alpha}\bigg) \right]
\end{equation}
which can be written as function of $p$ using Eqs.  (\ref{energiaC}) and (\ref{mu2})
or as a function of $\epsilon$ using Eqs.  (\ref{PC}) and (\ref{mu2B}).

In order to be absolutely stable, the energy per baryon of CFL quark matter  must be lower than the neutron
mass $m_n$ at $P=0$ and $T=0$ \cite{farhijaffe}.   Therefore,  we must have \cite{lugones2002}
\begin{equation}
\frac{\varepsilon}{ n_B} \bigg|_{P = 0} =  3 \mu   \leq  m_n = 939 \mathrm{MeV}.
\end{equation}
This  simple result is a direct consequence of the existence of a common Fermi momentum
for the three flavors in CFL quark matter and is valid at $T = 0$ without any approximation.
Since this must hold at the zero pressure point, then, from Eq. (\ref{mu2}) we have
\begin{equation}
B <  - \frac{m_s^2 m_n^2}{12 \pi^2} + \frac{\Delta^2 m_n^2}{3 \pi^2}
+  \frac{m_n^4}{108 \pi^2}
\label{windowanalytic}
\end{equation}
The last equation defines a region in the $m_s - B$ plane on which
the energy per baryon is smaller than $ m_n$ for a given $\Delta$.

Additionally, the empirically known stability of normal nuclear matter implies that the energy per baryon of a pure gas of quarks $u$ and $d$ at zero pressure and temperature must be higher than the neutron mass value.  Within the MIT bag model,  the latter condition imposes that $B > 57$  MeV  fm$^{-3}$ \cite{farhijaffe}.

The two above conditions, define the so called \textit{stability windows} shown in Fig. \ref{window} .  For a given value of $\Delta$,  Eq. (\ref{window}) gives the right side boundary of the window while the left side boundary is given
by the minimum value $B = 57$  MeV  fm$^{-3}$. As shown in \cite{lugones2002}, the window is significantly enlarged for increasing values of $\Delta$. This is to be compared, for example, with Fig. 1 of Ref. \cite{farhijaffe} in which no pairing was included.

Since the values of  $B$, $m_s$ and $\Delta$ are not accurately known we shall consider
them as free parameters in the equation of state. In Table \ref{models} we present all the sets of parameters  used in the present work and in Fig. \ref{window} we show how these  parametrizations are located inside the stability window.  Additionally, the parameters satisfy the stability condition $m_s^2 < 2 \mu \Delta$ given in Ref. \cite{alford2004}.

\begin{figure}[tb]
\includegraphics[angle=270,scale=0.35]{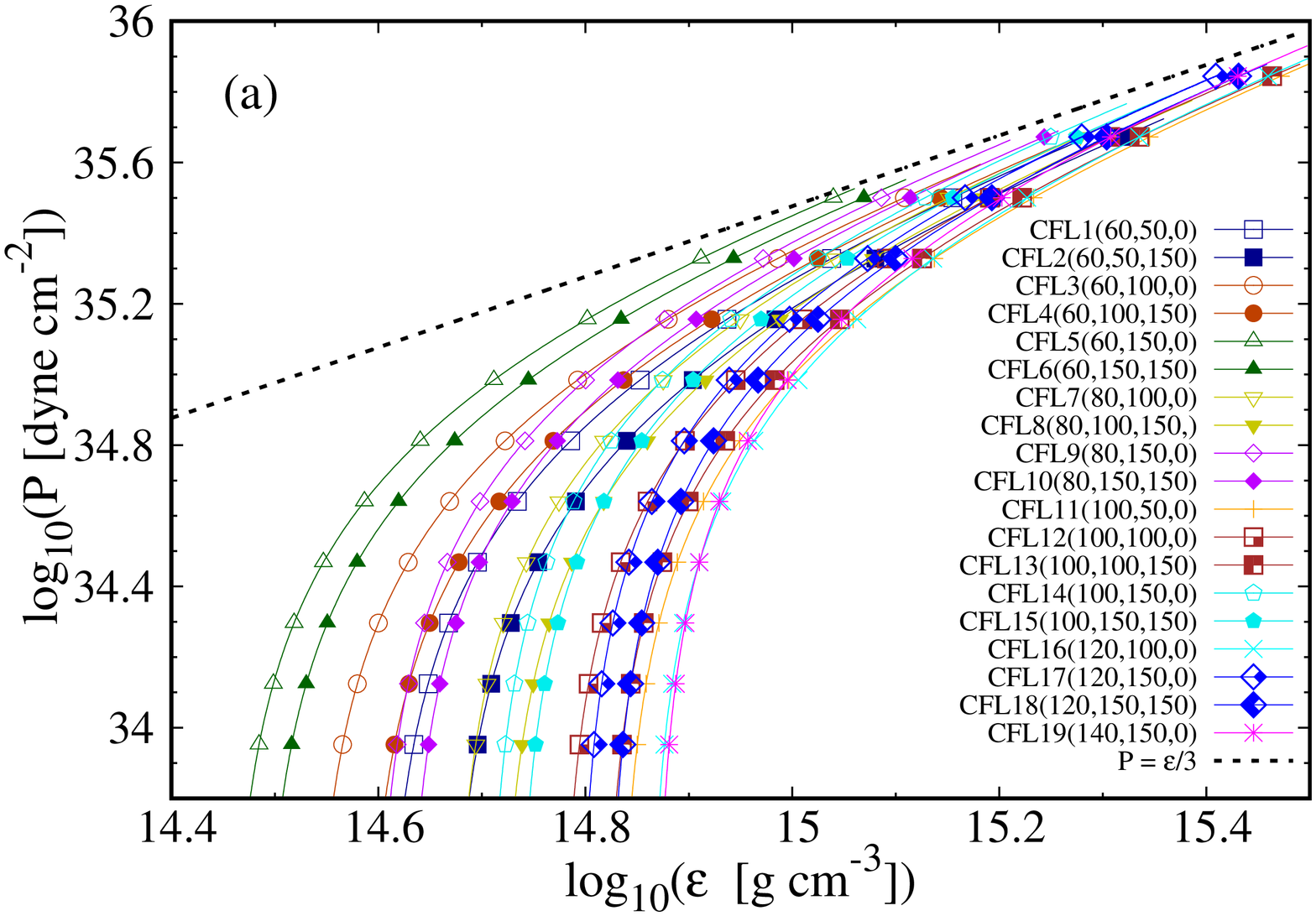}
\includegraphics[angle=270,scale=0.35]{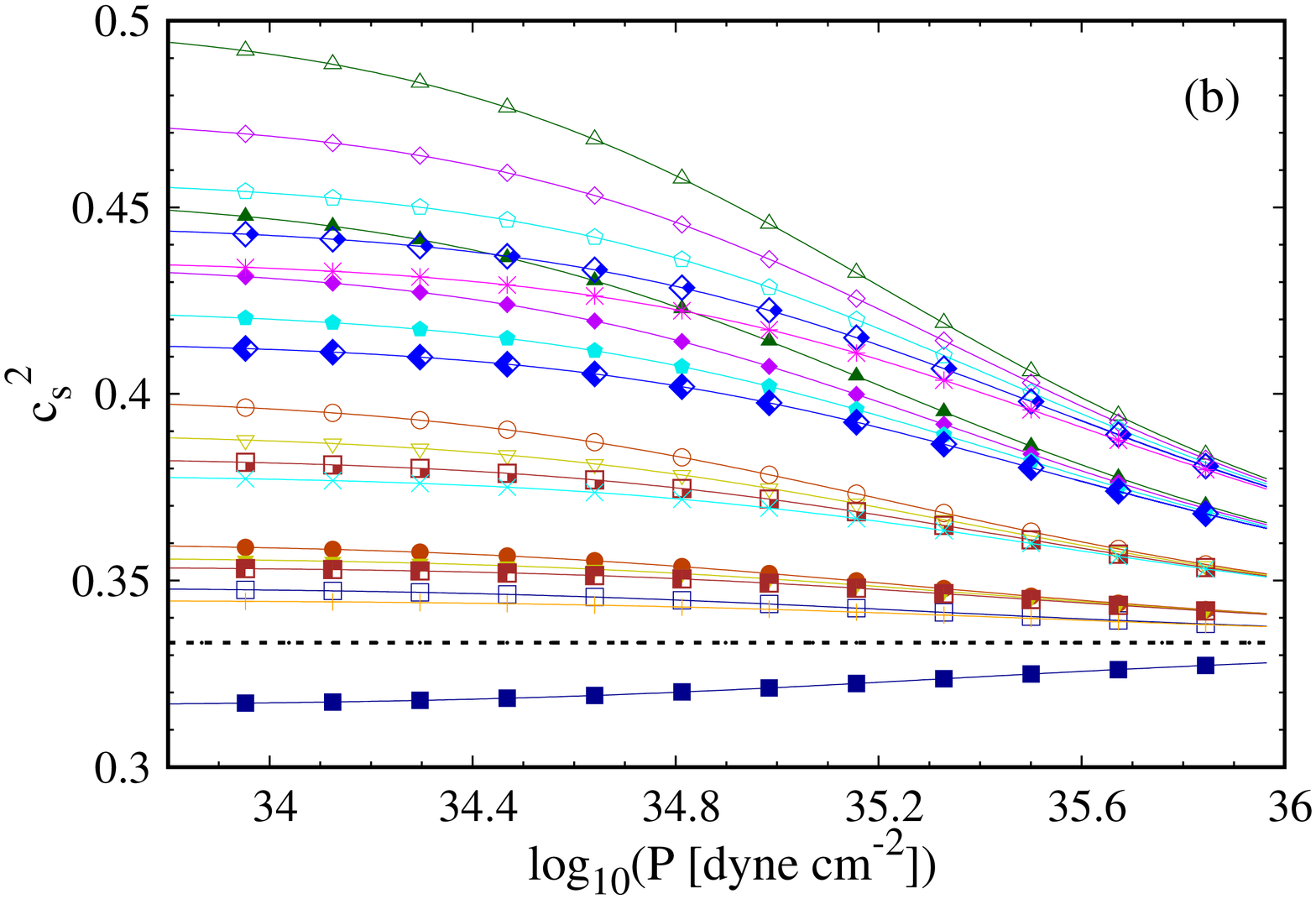}
\caption{The equation of state of CFL strange quark matter for all the parametrizations presented in Table \ref{models}. The slashed line corresponds to the ultrarelativistic limit $p=\epsilon/3$.}
\label{EOS}
\end{figure}

In Fig. \ref{EOS} we plot the pressure versus the energy density and the square of the speed of sound as a function of the pressure. As seen in Eq. (\ref{PC}), the pressure has a linear contribution $\epsilon/3$ that dominates asymptotically, but the term proportional to $\mu^2$ has a non negligible effect, leading to the non-linear behavior of the EOS apparent in Fig. \ref{EOS}. As a consequence, the speed of sound $c_s = \sqrt{dp/d\epsilon}$ tends to the ultra-relativistic limit $1/\sqrt{3}$ asymptotically but in general it has a density dependent value rather different from $1/\sqrt{3}$ \footnote{Constant-sound-speed parametrizations of the quark matter EOS can give useful insights for hybrid stars because only the high density regime where the EOS tends to be linear is relevant in those cases \cite{Zdunik2013,Chamel2013,Alford2015,Ranea-Sandoval2016}. For strange quark stars such approximation is less accurate.}.

\section{Non-radial oscillation equations of CFL stars}
\label{oscillation}

\subsection{Equilibrium configuration}

\begin{table}[tb]
\centering
\begin{tabular}{c | cccc}
\hline \hline
Model &      $\rho_{c}$      &       M$_{max}$      &          R      &         Z    \\
  &       [g cm$^{-3}$]      &       [M$_{\odot}$]      &           [km]     &           \\
\hline  
CFL1  \; & \;  1.902$\times 10^{15}$   \; & \;  2.051       \; & \;  11.08       	\; & \;      0.4849  \\
CFL2  \; & \;  2.286$\times 10^{15}$   \; & \;  1.830       \; & \;  10.09       	\; & \;      0.4673   \\ 
CFL3  \; & \;  1.520$\times 10^{15}$   \; & \;  2.357       \; & \;  12.38       	\; & \;      0.5111   \\   
CFL4  \; & \;  1.798$\times 10^{15}$   \; & \;  2.127     	 \; & \;   11.41       	\; & \;      0.4915  \\    
CFL5  \; & \;  1.148$\times 10^{15}$   \; & \;  2.842       \; & \;  14.24       	\; & \;      0.5603  \\    
CFL6  \; & \;  1.287$\times 10^{15}$   \; & \;  2.631       \; & \;  13.46       	\; & \;      0.5380  \\    
CFL7  \; & \;  2.110$\times 10^{15}$   \; & \;  1.994       \; & \;  10.52       	\; & \;      0.5065   \\  
CFL8  \; & \;  2.434$\times 10^{15}$   \; & \;  1.821       \; & \;   9.79       	\; & \;      0.4892  \\    
CFL9  \; & \;  1.623$\times 10^{15}$   \; & \;  2.365       \; & \;  11.98       	\; & \;      0.5484  \\   
CFL10 \; & \;  1.807$\times 10^{15}$   \; & \;  2.202       \; & \;  11.36       	\; & \;      0.5291  \\    
CFL11 \; & \;  3.204$\times 10^{15}$   \; & \;  1.571       \; & \;  8.51       	\; & \;      0.4824     \\
CFL12 \; & \;  2.701$\times 10^{15}$   \; & \;  1.754       \; & \;  9.29       	\; & \;      0.5030  \\   
CFL13 \; & \;  3.095$\times 10^{15}$   \; & \;  1.616       \; & \;  8.70       	\; & \;      0.4882  \\   
CFL14 \; & \;  2.105$\times 10^{15}$   \; & \;  2.055       \; & \;  10.49       	\; & \;      0.5395  \\    
CFL15 \; & \;  2.330$\times 10^{15}$   \; & \;  1.922       \; & \;  9.98       	\; & \;      0.5224  \\   
CFL16 \; & \;  3.311$\times 10^{15}$   \; & \;  1.582       \; & \;  8.40       	\; & \;      0.5007    \\ 
CFL17 \; & \;  2.606$\times 10^{15}$   \; & \;  1.834       \; & \;  9.42       	\; & \;      0.5332  \\  
CFL18 \; & \;  2.873$\times 10^{15}$   \; & \;  1.722       \; & \;  8.98       	\; & \;      0.5179  \\  
CFL19 \; & \;  3.145$\times 10^{15}$   \; & \;  1.667       \; & \;  8.60       	\; & \;      0.5293  \\
\hline \hline 
\end{tabular}
\caption{For each parametrization of the EOS we show the central energy density, the  mass, the radius and the surface gravitational redshift of the most massive object.} \label{maximum}
\end{table}

In order to study the radial oscillations of a compact star, we must determine first its equilibrium configuration.
Assuming that the unperturbed compact star is totally composed of a perfect fluid,  the stress-energy momentum tensor can be expressed as
\begin{equation}
{T}_{\mu\nu} =(\epsilon + p) u_{\mu}u_{\nu} + p{g}_{\mu\nu}.
\end{equation}

The generic background space-time of a static spherical star is expressed through the line element
\begin{equation}
\label{dsz_tov}
ds^{2}=-e^{ \nu(r)} dt^{2} + e^{ \lambda(r)} dr^{2} + r^{2}(d\theta^{2}+\sin^{2}{\theta}d\phi^{2}) .
\end{equation}
where $t, r, \theta, \phi$ are the set of Schwarzschild-like coordinates, and the metric potentials $\nu(r)$ and $\lambda(r)$  are functions of the radial coordinate $r$ only.

The Einstein equations in such a spacetime lead to the following set of stellar structure
equations (Tolman-Oppenheimer-Volkoff equations)
\begin{eqnarray}
\label{tov1}
&&\frac{dp}{dr} = - \frac{\epsilon m}{r^2}\bigg(1 + \frac{p}{\epsilon}\bigg)
	\bigg(1 + \frac{4\pi p r^3}{m}\bigg)\bigg(1 -
	\frac{2m}{r}\bigg)^{-1},
\\ \nonumber \\
\label{tov2}
&&\frac{d\nu}{dr} = - \frac{2}{\epsilon} \frac{dp}{dr}
	\bigg(1 + \frac{p}{\epsilon}\bigg)^{-1},
\\ \nonumber \\
\label{tov3}
&&\frac{dm}{dr} = 4 \pi r^2 \epsilon,
\end{eqnarray}
where $m$ is the gravitational mass inside the radius $r$. The metric function $\nu$ has the boundary condition
\begin{equation}
\label{BoundaryConditionMetricFunction}
    \nu(r=R)= \ln \bigg( 1-\frac{2M}{R} \bigg),
\end{equation}
where $R$ is the radius of the star and $M$ its mass. With this condition the  metric function $\nu(r)$ will match smoothly to the Schwarzschild metric outside the star.

\begin{figure*}[tb]
\centering
\includegraphics[angle=270,scale=0.35]{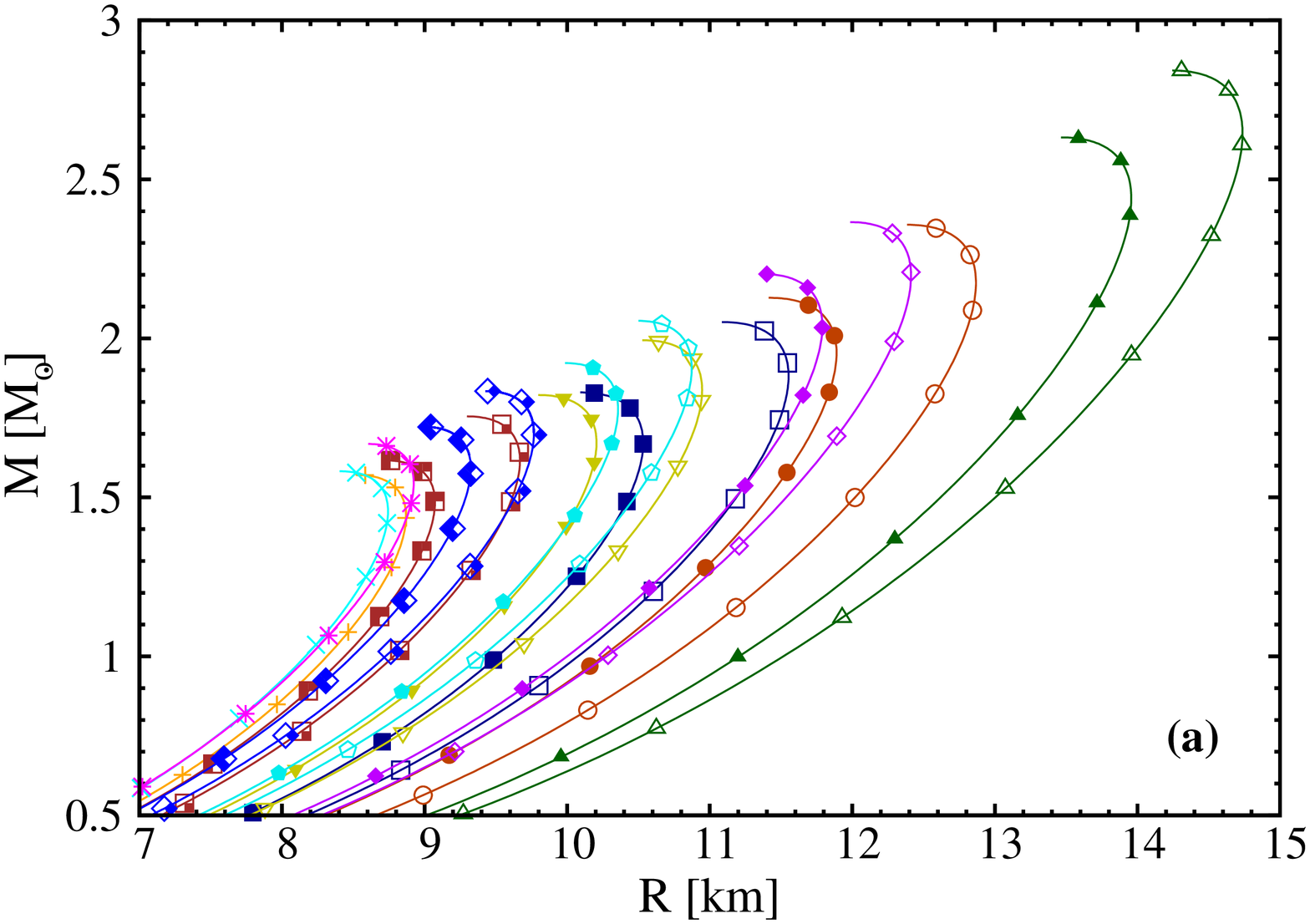} 
\includegraphics[angle=270,scale=0.35]{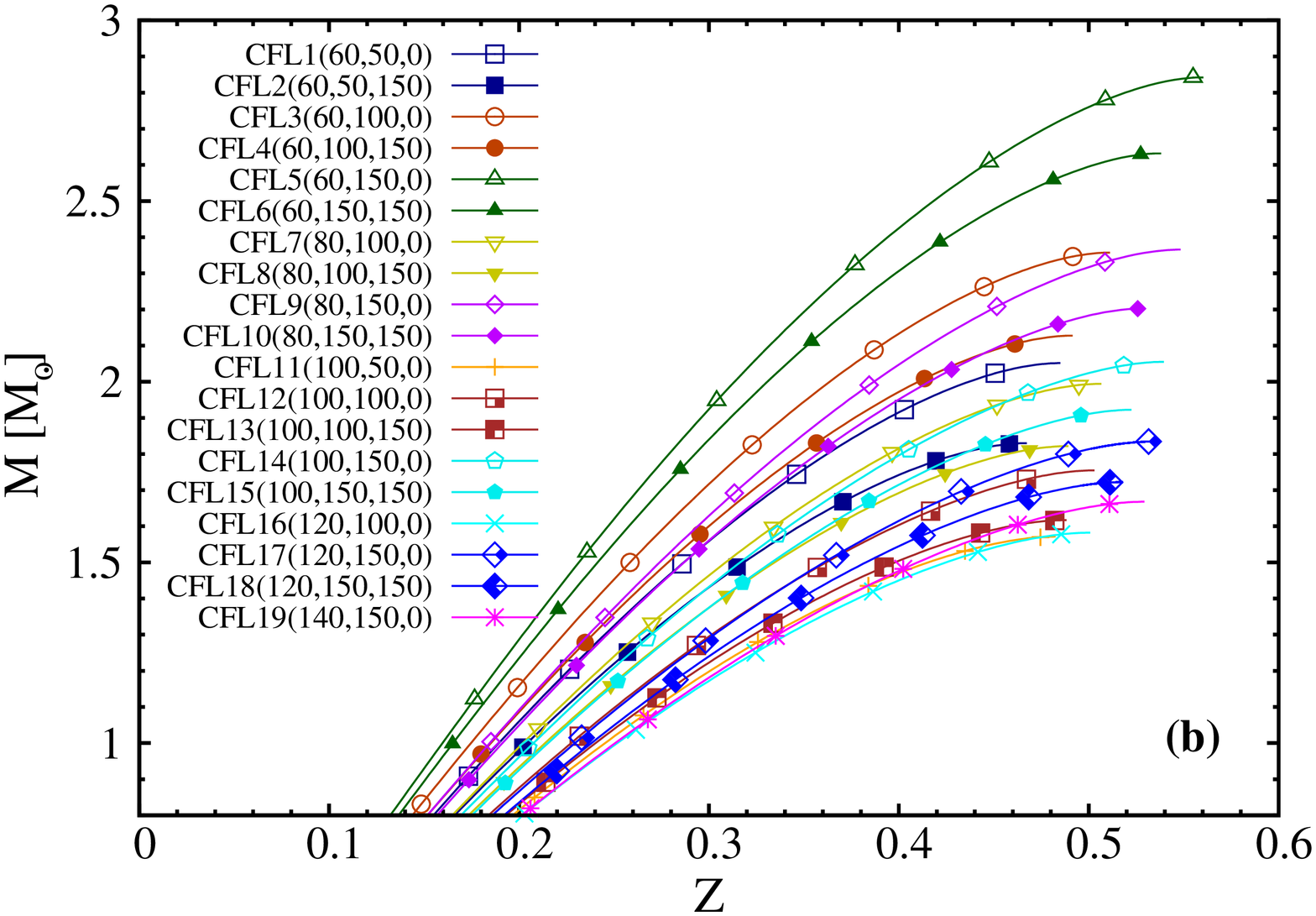}
\caption{Mass of CFL stars as a function of the radius $R$ and the surface gravitational redshift $Z$.}
\label{stellar_structure}
\end{figure*}

In Table \ref{maximum} we present the properties of the maximum mass objects for each parametrization of the equation of state. 
The parameters were chosen in order to cover  the entire stability windows. However,  values to the right in Fig. \ref{window}, e.g. models CFL16, CFL17, CFL18 and CFL19 result in small maximum masses (see Table \ref{stellar_structure}). In fact, since the maximum mass decreases with $B$ (see \cite{lugones2004}), models to the right of CFL19 in Fig. \ref{window}  are incompatible with the existence of the massive pulsars PSR J1614-2230 with $M = (1.97 \pm 0.04) M_{\odot}$ \cite{demorest2010} and PSR J0348-0432 with $M = (2.01 \pm 0.04) M_{\odot}$ \cite{antoniadis2013}.

\subsection{Non-radial oscillation equations}

To obtain the equations that govern the non-radial oscillations, both
fluid and spacetime variables are perturbed. These perturbations
are inserted into the Einstein equations and into the energy, momentum and
baryon number conservation equations and only the first-order terms are retained.

We use the Lindblom-Detweiler ansatz \cite{lindblom1983,detweiler1985} for the polar perturbations in the metric tensor 
\begin{eqnarray}
ds^2 & = & -e^{\nu}(1+r^{\ell}H_0Y^{\ell}_{m}e^{i\omega t})dt^2 \nonumber \\
&& - 2i\omega r^{\ell+1}H_1Y^{\ell}_me^{i\omega t}dtdr +  \nonumber \\
&& + e^{\lambda}(1 - r^{\ell}H_0Y^{\ell}_{m}e^{i \omega t})dr^2 \nonumber \\
&& + r^2(1 - r^{\ell}KY^{\ell}_{m}e^{i \omega t})(d\theta^2 + \sin^2\theta d\phi^2). 
\end{eqnarray}
and consider the polar perturbations in the fluid described by the following Lagrangian displacements
\begin{eqnarray}
\xi^{r} &=& r^{\ell-1}e^{-\lambda/2}WY^{\ell}_{m}e^{i\omega t}, \\
\xi^{\theta} &=& -r^{\ell - 2}V\partial_{\theta}Y^{\ell}_{m}e^{i\omega t}, \\
\xi^{\phi} &=& -r^{\ell}(r \sin \theta)^{-2}V\partial_{\phi}Y^{\ell}_{m}e^{i\omega t} . 
\end{eqnarray}

With this, non-radial oscillations are described by the following first order linear system of differential equations \cite{detweiler1985}:
\begin{widetext}
\begin{eqnarray}
H_1' &=&  -r^{-1} \biggl[ \ell+1+ 2Me^{\lambda} r^{-1}+4\pi   r^2 e^{\lambda}(p-\epsilon) \biggr]H_1
 +  e^{\lambda}r^{-1}  \left[ H_0 + K - 16\pi(\epsilon+p)V \right] \:,      \label{osc_eq_1}  \\
 K' &=&    r^{-1} H_0 + \frac{\ell(\ell+1)}{2}r^{-1} H_1   - \left[(\ell+1)r^{-1}-\frac{\nu'}{2} \right] K
 - 8\pi(\epsilon+p) e^{\lambda/2}r^{-1} W \:,  \label{osc_eq_2} \\
 W' &=&  - (\ell+1)r^{-1} W   + re^{\lambda/2} \left[ e^{-\nu/2} \gamma^{-1}p^{-1}X
 - \ell(\ell+1)r^{-2} V + \frac{1}{2}H_0 + K \right] \:,  \label{osc_eq_3}\\
 X' &=&  - \ell r^{-1} X + \frac{(\epsilon+p)e^{\nu/2}}{2}  \Biggl[ \left( r^{-1}+\frac{\nu'}{2} \right)H_0
 + \left(r\omega^2e^{-\nu} + \frac{\ell(\ell+1)}{2}r^{-1}\right) H_1   + \left(\frac{3}{2}\nu' - r^{-1}\right) K \nonumber  \\
&& - \ell(\ell+1)r^{-2}\nu' V 
 -  2 r^{-1} 
 \Biggl( 4\pi(\epsilon+p)e^{\lambda/2} + \omega^2e^{\lambda/2-\nu}
 - \frac{r^2}{2}
 \biggl(e^{-\lambda/2}r^{-2}\nu'\biggr)' \Biggr) W \Biggr]  \:,
\label{osc_eq_4}
\end{eqnarray}
where the prime denotes a derivative with respect to $r$, $\gamma$ is the adiabatic index,  $X$ is given by
\begin{equation}
X = \omega^2(\epsilon+p)e^{-\nu/2}V - \frac{p'}{r}e^{(\nu-\lambda)/2}W + \frac{1}{2}(\epsilon+p)e^{\nu/2}H_0 ,
\end{equation}
and $H_{0}$ fulfills the algebraic expression 
\begin{eqnarray}
\left[ 3M + \frac{1}{2}(l+2)(l-1)r + 4\pi r^{3}p\right] H_{0}=8\pi r^{3}e^{-\nu /2}X - \left[ \frac{1}{2}l(l+1)(M+4\pi r^{3}p)-\omega^2
r^{3}e^{-(\lambda+\nu)}\right] H_{1}\nonumber \\
+ \left[ \frac{1}{2}(l+2)(l-1)r - \omega^{2} r^{3}e^{-\nu}-r^{-1}e^{\lambda}(M+4\pi r^{3}p)(3M - r + 4\pi r^{3}p)\right] K.
\end{eqnarray}

\end{widetext}

The above system of equations is used to describe the behavior of perturbations inside the star ($0 < r < R$).
Outside of the star ($r >R$), the perturbation functions that describe the motion of the fluid vanish and the above system of equations reduces to a second order differential equation, namely the Zerilli equation:
\begin{equation}
\frac{d^{2}Z}{dr^{*2}}=[V_{Z}(r^{*})-\omega^{2}]Z,
\end{equation}
where the Zerilli function $Z(r^{*})$ and its derivative $dZ(r^{*})/dr^{*}$  are related to the metric perturbations  $H_{0}(r)$ and $K(r)$ by the transformation given in Eqs. (A27)$-$(A34) of Ref. \cite{lindblom1983} (for the correction of a typographical error in Eq. (A29), see \cite{detweiler1985}).  $Z(r^{*})$  depends on the ``tortoise'' coordinate $r^{*} = r + 2 M \ln (r/ (2M) -1)$ and  the effective potential  $V_{Z}(r^{*})$ is given by
\begin{eqnarray}
V_{Z}(r^{*}) = \frac{(1-2M/r)}{r^{3}(nr + 3M)^{2}}[2n^{2}(n+1)r^{3} \nonumber \\
+ 6n^{2}Mr^{2}+18nM^{2}r + 18M^{3}],
\end{eqnarray}
being $n= (l-1) (l+2) / 2$.

\subsection{Boundary conditions}

The system of Eqs. (\ref{osc_eq_1})$-$(\ref{osc_eq_4}) has four linearly independent solutions for given values of $l$ and $\omega$.  The physical solution needs to verify the appropriate boundary conditions:  (a)  The perturbation functions must be finite everywhere, in particular at $r = 0$ where non-radial oscillation equations are singular. To implement such condition  it is necessary to use a power series expansion of the solution near the singular point $r=0$ (the procedure is explained in detail in \cite{lindblom1983}).
(b) The Lagrangian perturbation in the pressure has to be zero at the surface of the star $r = R$. This implies that the function $X$ must vanish at $r = R$.  Given specific values $l$ and $\omega$,   there is a unique solution that satisfies the above boundary conditions inside the star.  

In general, outside the star the perturbed metric describes a combination of outgoing and ingoing gravitational waves.  The physical solution of the Zerilli equation is the one that describes purely outgoing gravitational radiation at $r=\infty$. 
Such boundary condition cannot be verified by any value of $\omega$ and the frequencies that fulfill this requirement represent the quasi normal modes of the stellar model.

\subsection{Numerical procedure} 

To calculate the frequency of the fundamental mode we employed the method described in \cite{lindblom1983}.   First we adopt an adequate trial value as an input for the frequency $\omega$.  We can use the Newtonian frequency  $\omega = \sqrt{(M/R^{3})2l(l-1)/(2l+1)}$ or  a frequency calculated within the Cowling approximation  (see e.g. \cite{vasquez2014}). 
With such guess we begin the integration inside the star. First, we select three linearly independent solutions compatible with the regularity conditions at the stellar center, and numerically integrate from $r=0$ to the point $R/2$ inside the star.  As a second step we select two linearly independent solutions  compatible with the boundary condition at the surface, and numerical integrate them from the surface $R$ to the point $R/2$. To complete the procedure inside the star, we combine the previous five solutions to obtain one which is compatible with the  boundary conditions at the center and at the surface.

Thereafter, we continue with the integration outside the star. The boundary values for the Zerilli function and its derivative  at the surface of the star  can always be obtained from the values of $H_{0}(R)$ and $K(R)$ arising from the inside integration. Then,  $Z(r^{*})$ can be determined outside the star by integrating the Zerilli equation.  At asymptotically large radius, the two linearly independent solutions of the Zerilli equation may be expressed as power series $Z_{-}(r^{*})=e^{-i\omega r^{*}}\sum_{j=0}^{\infty}\beta_{j}r^{-j}$ and $Z_{+}(r^{*})=e^{i\omega r^{*}}\sum_{j=0}^{\infty}\overline{\beta}_j r^{-j}$,  where $Z_{-}$ represents the outgoing waves and  $Z_{+}$ the ingoing waves.  The expressions for the coefficients $\beta_{j}$ and their complex conjugate $\overline{\beta}_{j}$ can be found in \cite{jun-li2011}. The general asymptotic solution is given by the linear combination 
\begin{equation}\label{linealzerilli}
Z(r^{*}) = A(\omega) Z_{-}(r^{*}) + B(\omega) Z_{+}(r^{*}),
\end{equation}
where $A(\omega)$ and  ${B}(\omega)$ are complex numbers.
The integration of the Zerilli equation from the surface of the star allows finding  $Z$ and $dZ/dr$  at infinity, more specifically  at  $r_{\infty} \approx 50 \, \omega^{-1}$. Then, using Eq. (\ref{linealzerilli}) at $r^* = r^*_{\infty}$, we can obtain $B(\omega)$ as a function  of $Z(r^*_{\infty})$ and $dZ/dr(r^*_{\infty})$ for each value of $\omega$ (which is treated as a real number when integrating Eqs. (\ref{osc_eq_1})$-$(\ref{osc_eq_4}) and the Zerilli equation). 
The boundary condition at infinity requires that we only have outgoing gravitational radiation, i.e. in order to find the frequencies of quasi normal modes we must obtain the roots of $B(\omega)=0$. This is achieved by first determining $B(\omega)$ for three different close values of  the real trial frequency $\omega$. Then, we fit a quadratic polynomial $B(\omega)= \gamma_0 + \gamma_1 \omega + \gamma_2 \omega^2$ to the computed values of $\omega$ and locate the complex roots of the polynomial expansion,  in order to obtain an approximate root of $B(\omega)=0$. Finally, we iterate this procedure using the real part of the approximate root as an input of the next integration of the oscillation equations. This process is repeated until the real part value of the quasi normal mode  changes from one step to the next by less than one part in $10^8$. The imaginary part of $\omega$ is related to the gravitational damping time of the oscillation, $\tau = 1 / \mathrm{Im}(\omega)$.

\section{Results}
\label{results}

\begin{figure*}[tp]
 \centering
\includegraphics[angle=270,scale=0.35]{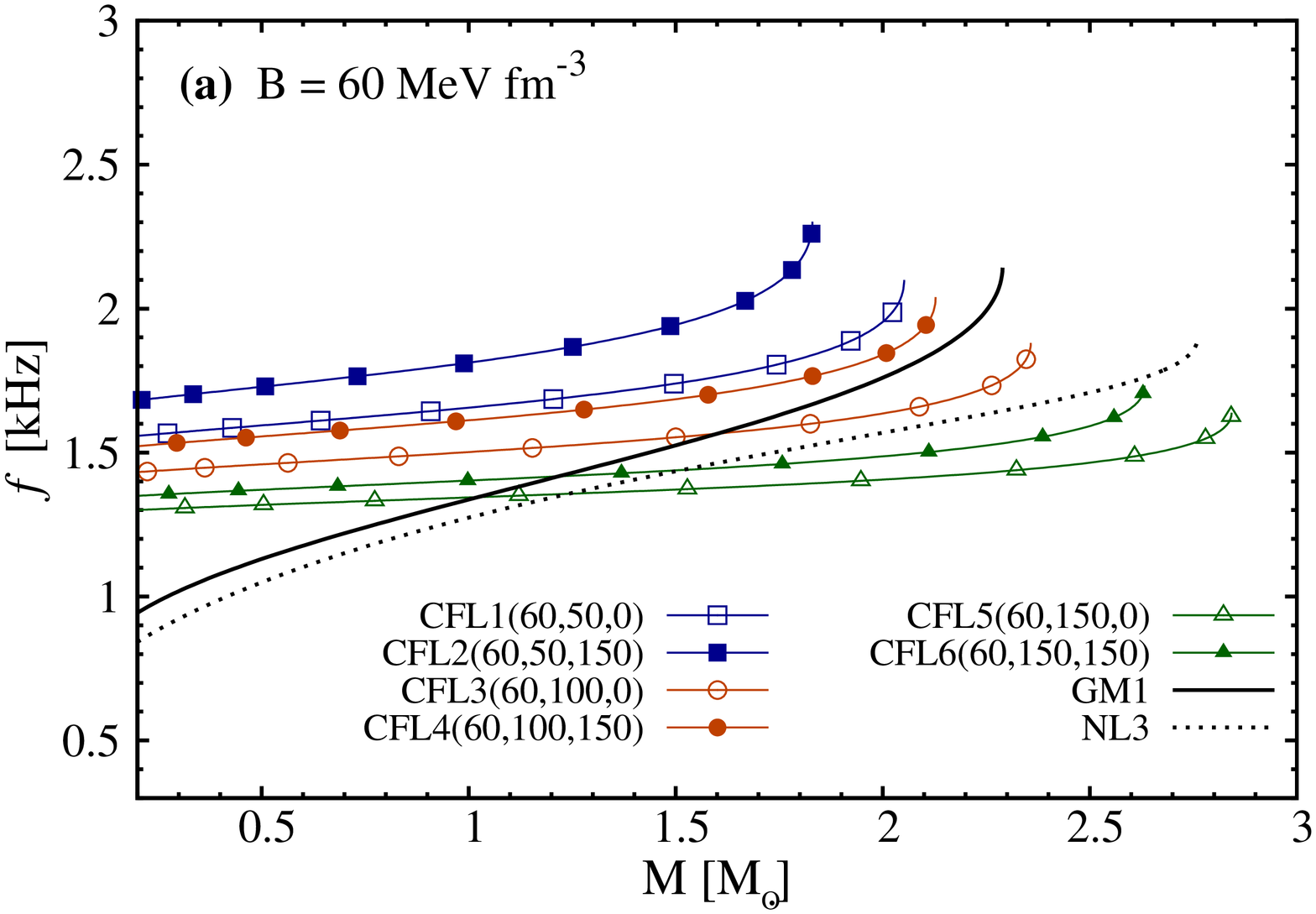}
\includegraphics[angle=270,scale=0.35]{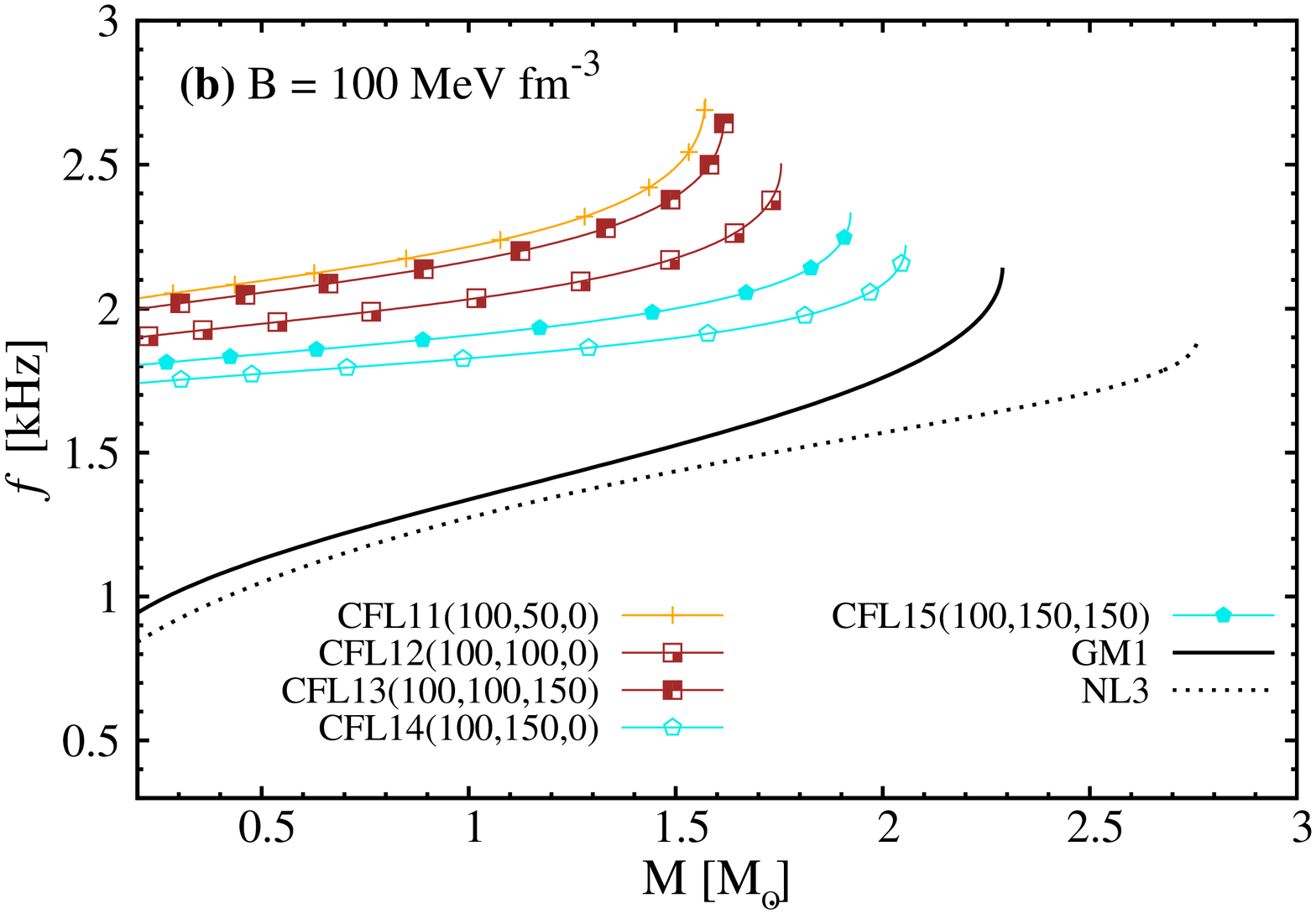}
\includegraphics[angle=270,scale=0.35]{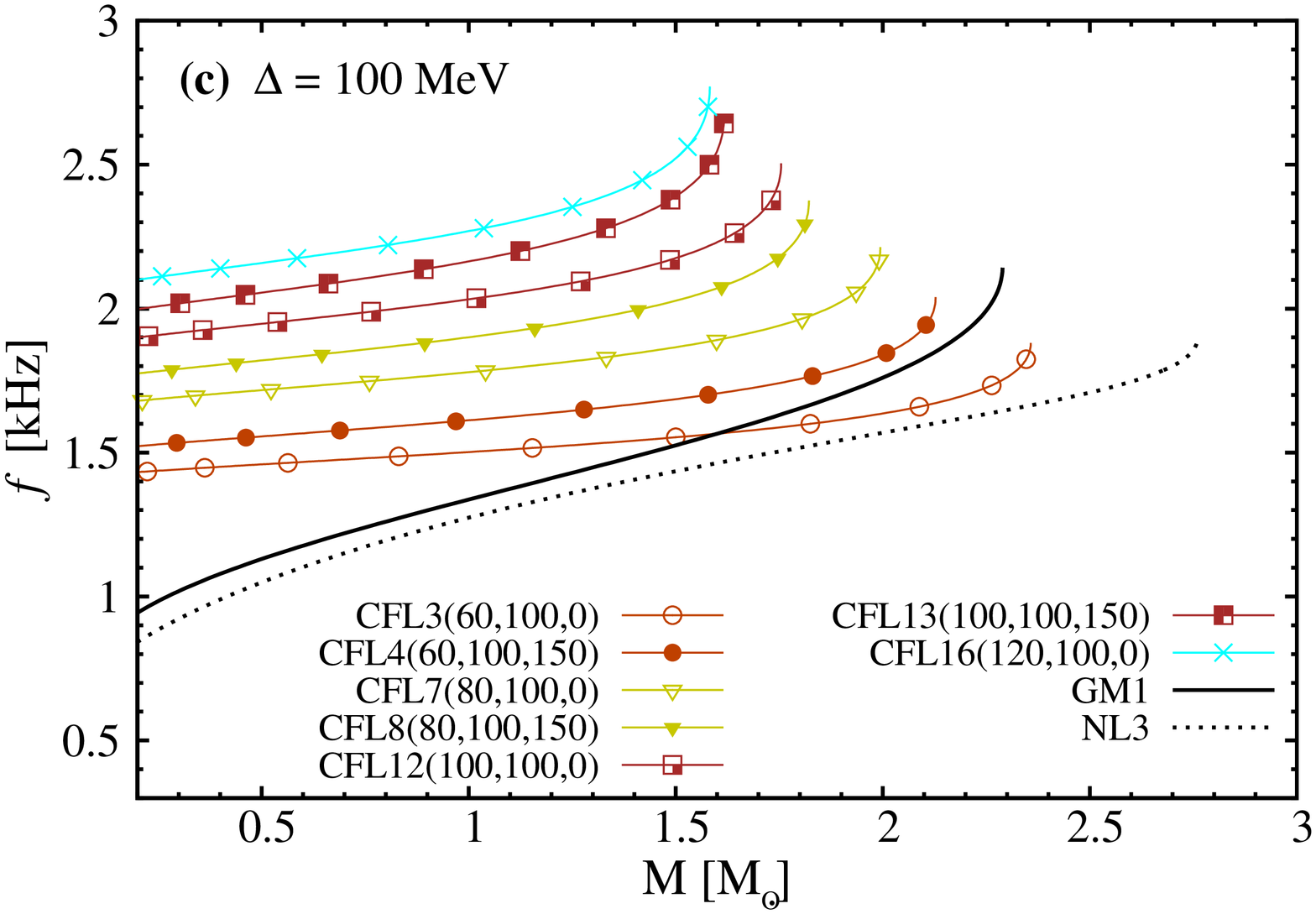}
\includegraphics[angle=270,scale=0.35]{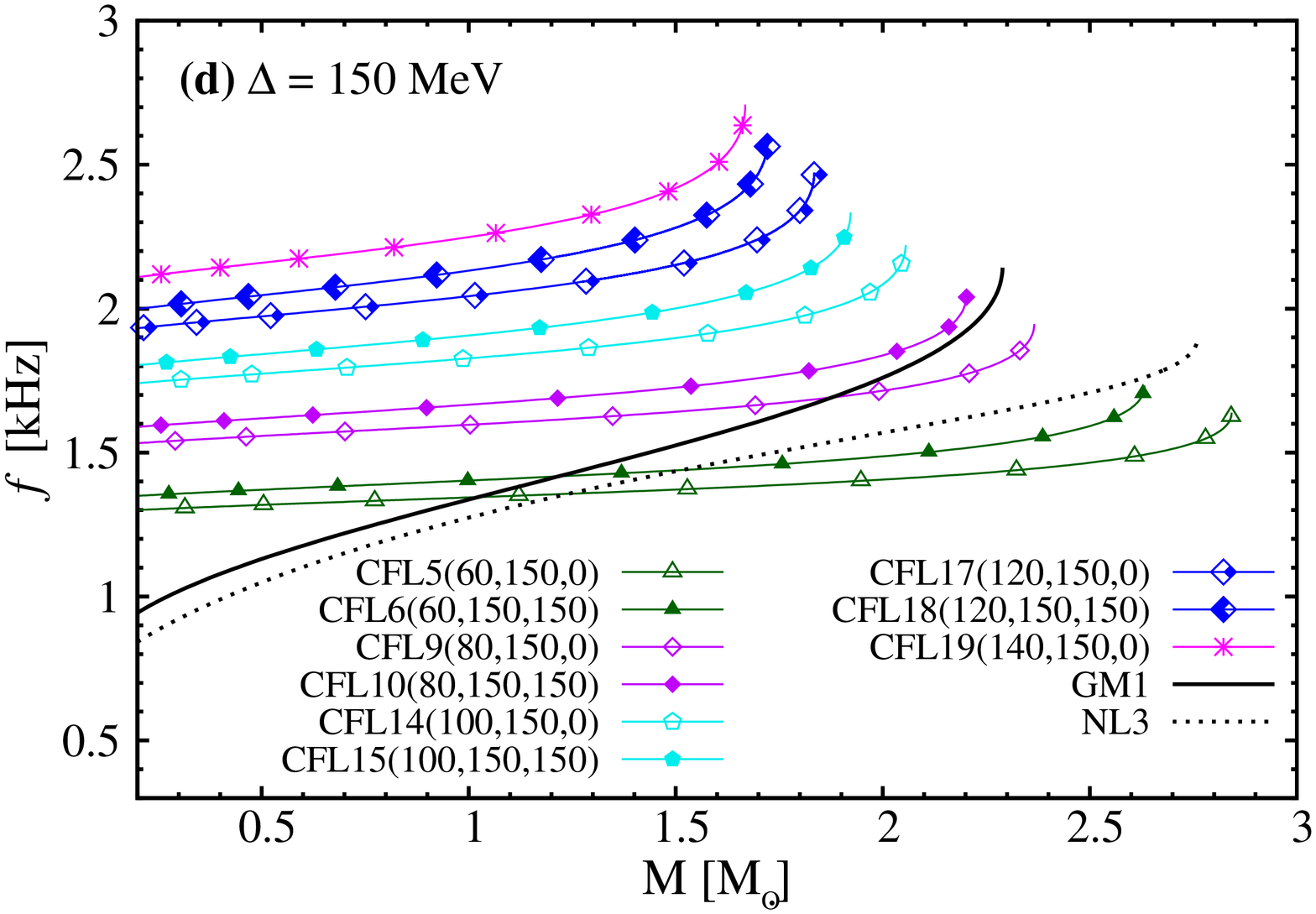}
\caption{Frequency of the fundamental oscillation mode  as a function of the stellar mass $M$. In panels (a) and (b) we fixed the bag constant $B$ to 60 and 100 MeV fm$^{-3}$ respectively, and varied the pairing gap $\Delta$. In panels (c) and (d) we fixed the $\Delta$ parameter  to 100 and 150 MeV, and varied the bag constant $B$. The curves are identified with the label introduced in Table \ref{models} followed by the value of the three parameters of the EOS: ($B$  [MeV fm$^{-3}$], $\Delta$  [MeV], $m_s $ [MeV]). For comparison we show the results for hadronic stars using the relativistic mean field model EOSs GM1 and NL3 with nucleons and electrons (see e.g. \cite{Lenzi2012} and references therein).}
\label{frequency}
\end{figure*}

\begin{figure*}[tp]
 \centering
\includegraphics[angle=270,scale=0.35]{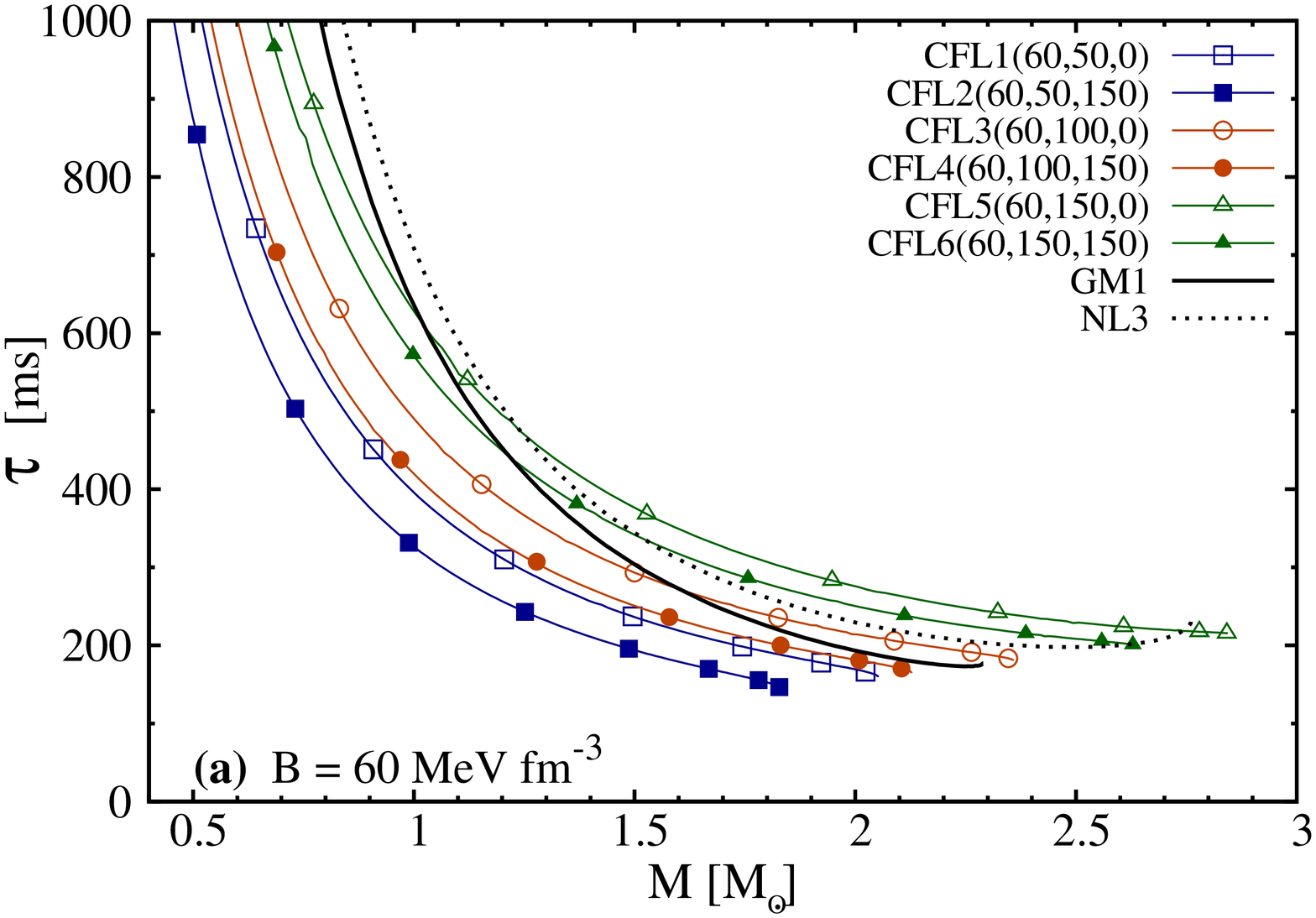}
\includegraphics[angle=270,scale=0.35]{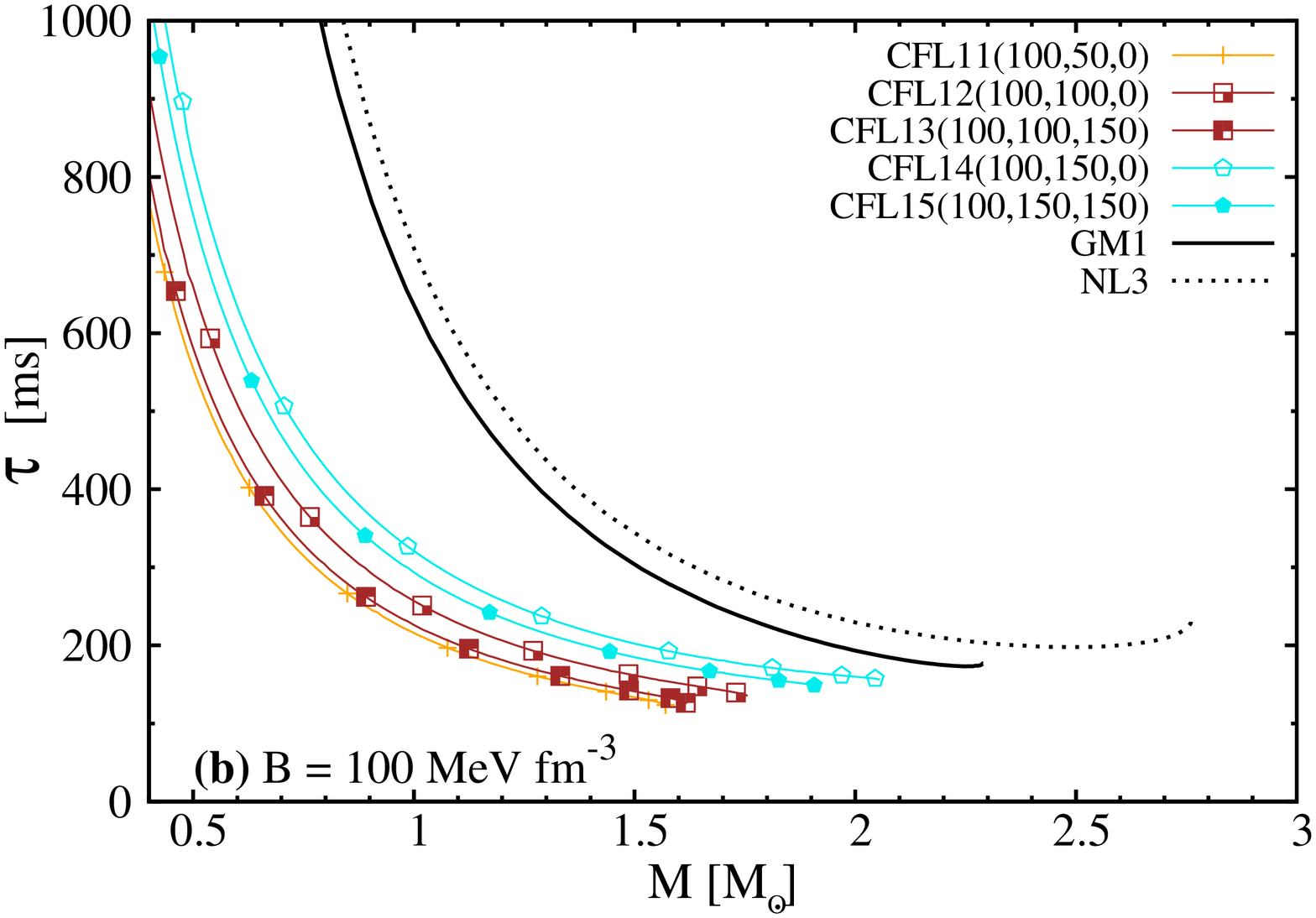}
\includegraphics[angle=270,scale=0.35]{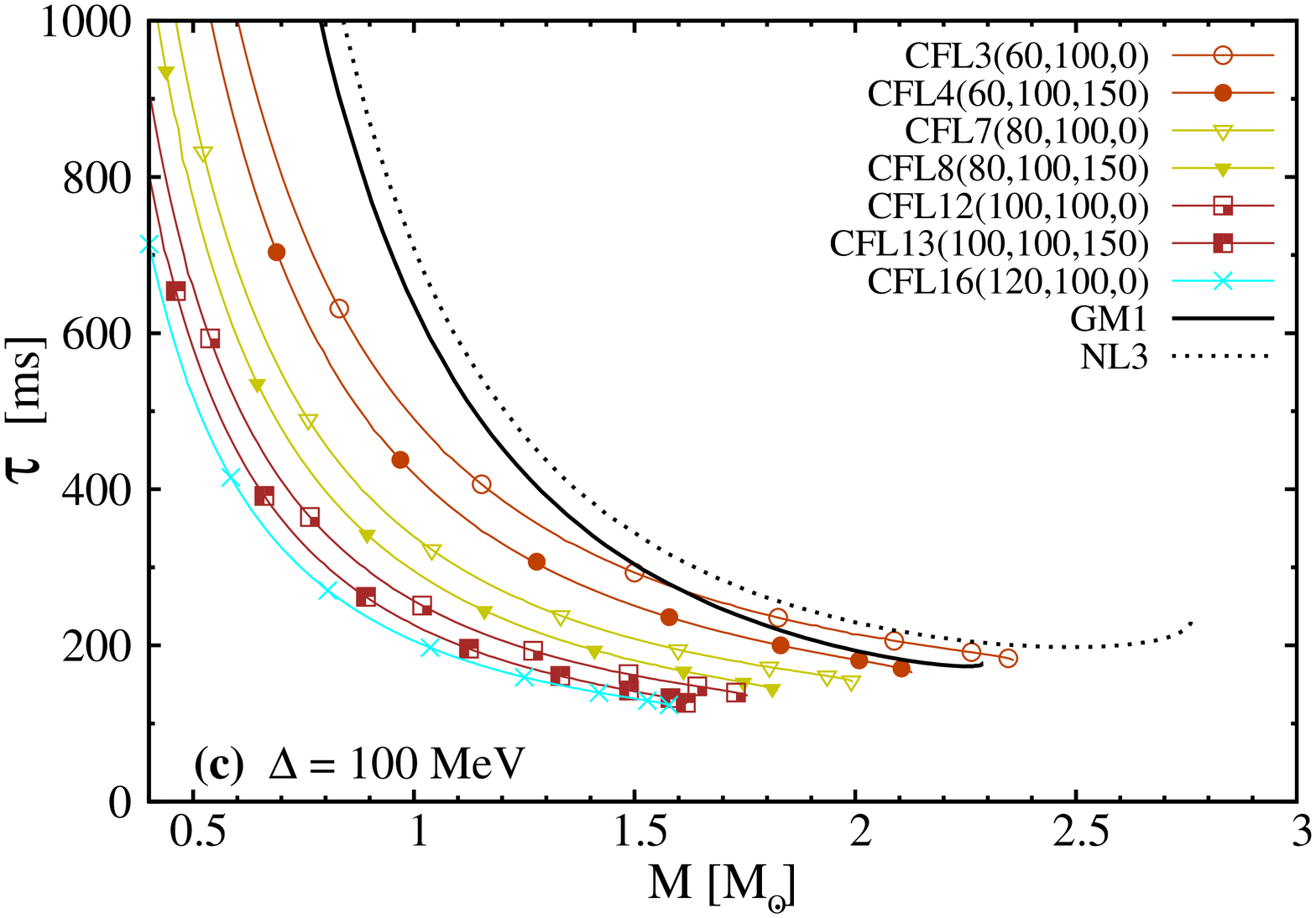}
\includegraphics[angle=270,scale=0.35]{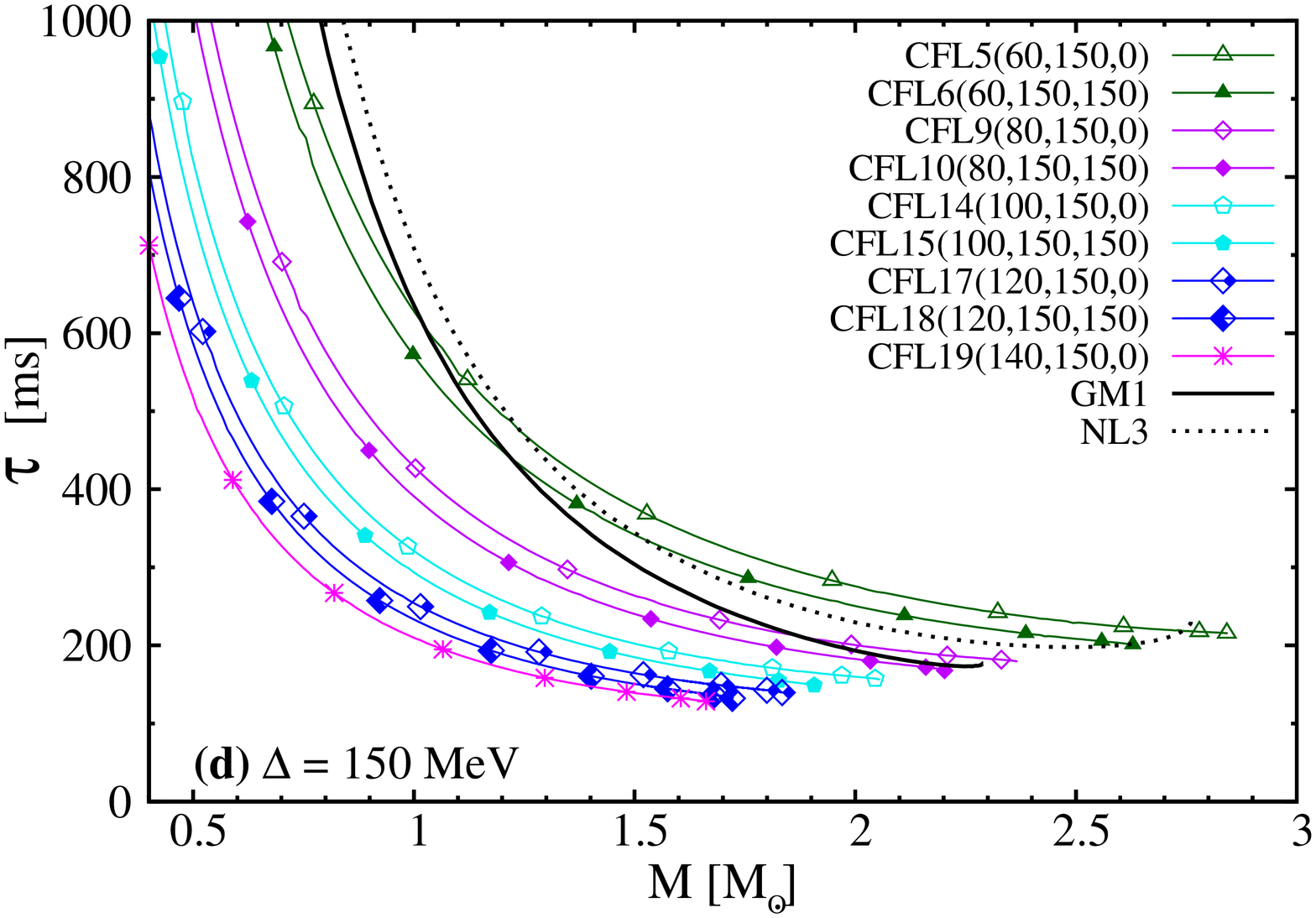}
\caption{The damping time of the fundamental oscillation mode as a function of the stellar mass $M$ for the same models shown in Fig. \ref{frequency}.}
\label{damping}
\end{figure*}

\begin{figure}[tb]
 \centering
\includegraphics[angle=270,scale=0.35]{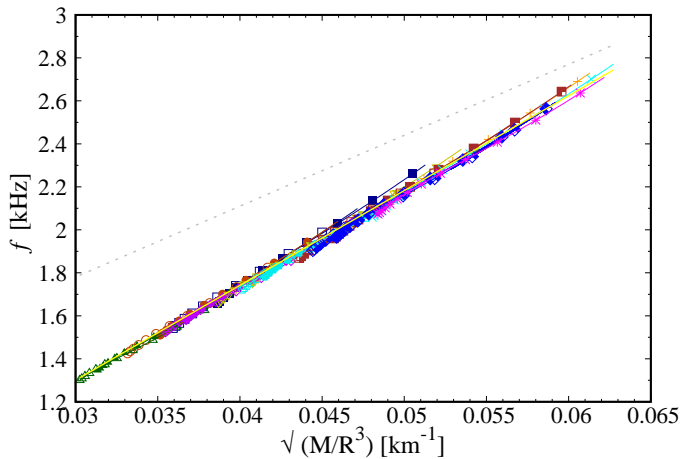}
\caption{The frequency of the fundamental mode as a function of the square root of the average density for the different parametrizations of the EOS considered in this paper. The full yellow line is the fitting given in Eq. \eqref{fitting_f} {and the dotted line is the fitting for hadronic stars given in Ref. \cite{benhar2004}.} }
\label{fig:linear_f}
\end{figure}

\begin{figure}[tbh]
 \centering
\includegraphics[angle=270,scale=0.35]{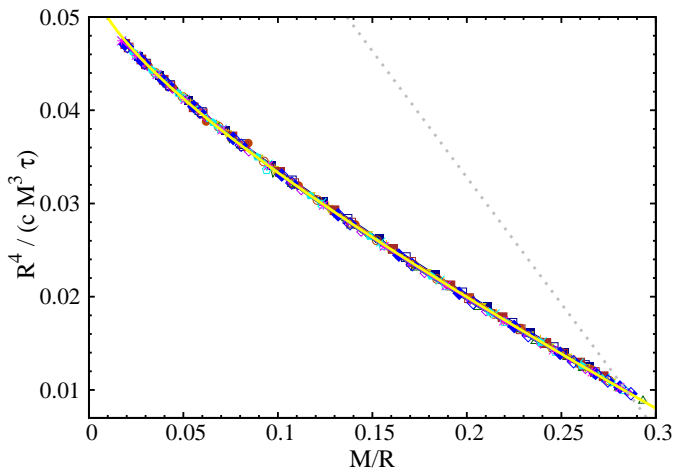}
\caption{The damping time of the fundamental mode as a function of the compactness $M/R$ for the different parametrizations of the EOS considered in this paper. The full yellow line is the fitting given in Eq. \eqref{fitting_tau} {and the dotted line is the fitting for hadronic stars given in Ref. \cite{benhar2004}.} }
\label{fig:linear_tau}
\end{figure}

\begin{figure}[tb]
 \centering
\includegraphics[angle=270,scale=0.35]{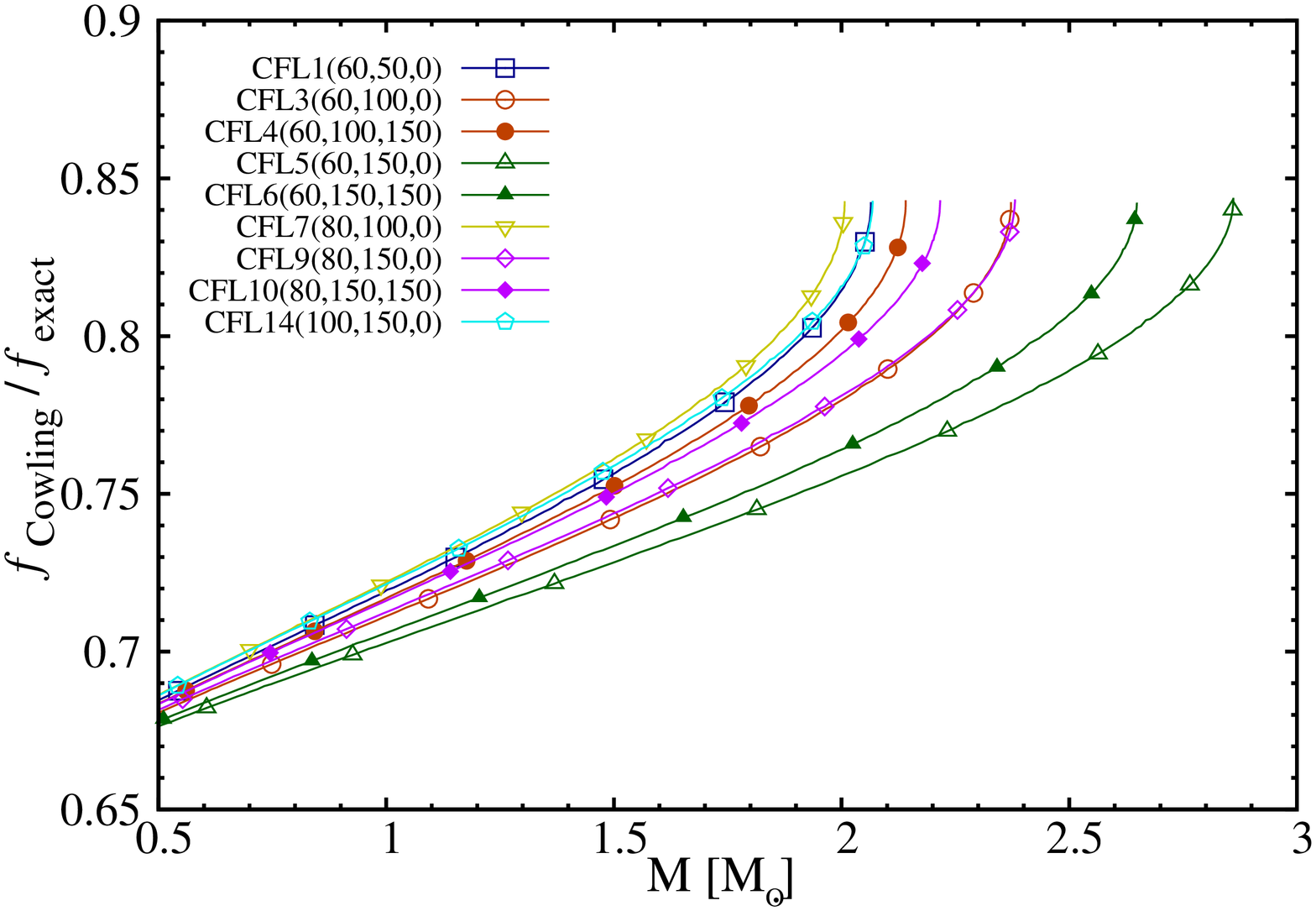}
\caption{Comparison with the Cowling approximation. The frequency  $f_{exact}$ is  obtained through the full calculations of the present paper while $f_{Cowling}$ is obtained within the Cowling approximation  in which the metric perturbations are neglected (see \cite{vasquez2014,vasquez2014nach}). }
\label{fig:cowling}
\end{figure}

In this section we obtain the frequency  $f = \mathrm{Re}(\omega)/ 2 \pi$ and the gravitational damping time $\tau = 1/ \mathrm{Im}(\omega)$  of the fundamental mode of pulsation of  CFL compact stars for different stellar configurations and for quadrupole oscillations ($l=2$).
Specifically we focus on how  $f$ and $\tau$ depend on  the compact star mass $M$. The dependence on the stellar radius $R$ and  the surface gravitational redshift $z$ can be derived with the help of Fig. \ref{stellar_structure}.  We also analyze the dependence of $f$ and  $\tau$ on the EOS's parameters $B$, $\Delta$ and $m_s$.

In Fig. \ref{frequency} we see that the oscillation frequency for the fundamental mode of CFL strange stars is always in the range $\sim 1-3$ kHz. For a given choice of the EOS's parameters, the oscillation frequency doesn't vary strongly with the stellar mass;  typically, a star with 0.5 $M_{\odot}$ and a star with a mass close to the maximum one, have frequencies that differ by less than $\sim 30 \%$. The dependence of $f$ on the EOS's parameters is also shown in Fig. \ref{frequency}. For larger $\Delta$ the maximum stellar mass is larger, and for larger  $B$ or $m_s$ it is smaller. This produces a right-left shifting of the curves when varying the EOS's parameters. Variations in the EOS's parameters  also produce an up-down shifting of the curves. As a result we find that: (i)  an increase in $\Delta$ keeping $B$ and $m_s$ fixed, shift the $f$ versus $M$ curves downwards and to the right; (ii) an increase in $B$ or $m_s$ keeping the other parameters fixed, shift the curves upwards and to the left.

In Fig. \ref{damping} we see that the damping time is in the range $\sim 0.2-1$ s for CFL strange stars with $M > 0.5 M_{\odot}$. We also see that if $\Delta$ is increased for fixed $B$ and $m_s$, there is a shift of the curves upwards and to the right (panels (a) and (b)). If $B$ (or $m_s$) is increased keeping the other parameters fixed, there is a shift of the curves downwards and to the left (panels (c) and (d)).

In Fig.  \ref{fig:linear_f} we show the $f$-mode frequency as a function of the square root of the average stellar density, $\sqrt{M/R^3}$, which is known to be the natural scaling of the mode. Notice that in spite of the very large degree of uncertainty in the choice of the EOS’s parameters,  the results for all parametrizations fall within a very narrow region and can be fitted by the same  linear empirical relation: 
\begin{equation}
f =  a_1 +    a_2  \sqrt{M/R^{3}} , 
\label{fitting_f}
\end{equation}
with  and $a_1 = - 0.023$  kHz  and   $a_2 = 44.11$  km/kHz.

In Fig.  \ref{fig:linear_tau} we show the damping time of the fundamental mode  as a function of the compactness $M/R$.  Again,  the results for all parametrizations can be fitted by the same  empirical relation: 
\begin{equation}
\left( \frac{M^3 \tau}{R^4} \right)^{-1} = b_1 + b_2   \sqrt{\frac{M}{R}} + b_3 \frac{M}{R} ,
\label{fitting_tau}
\end{equation}
with $b_1 = 0.0553$,  $b_2 =-0.0466$  and $b_3 = - 0.0725$. 

As emphasized in \cite{andersson1998,andersson1999,benhar2004}  these two empirical relations could be used to determine the mass and the radius of the star from the knowledge of the frequency and damping time of the modes.  Moreover, such method looks particularly promising in the case of CFL strange stars, because as shown in Figs. \ref{fig:linear_f} and \ref{fig:linear_tau} the results for all parametrizations of the EOS fall within a very narrow region around the fitting curves. {For comparison, we also show in Figs. \ref{fig:linear_f} and \ref{fig:linear_tau} the fittings presented in Ref. \cite{benhar2004} for hadronic stars, which were obtained using the EOSs of  APR \cite{APR} and GM \cite{GM} for the stellar core.  The curves for hadronic stars and CFL strange stars are very different, meaning that these results can potentially help in the discrimination of the internal composition of compact objects. }

Finally, we have compared our results with previous calculations of the oscillation of CFL strange stars  made within the relativistic Cowling approximation, where the metric perturbations are set to zero \cite{vasquez2014,vasquez2014nach}. As shown in Fig. \ref{fig:cowling}, for  stars with $M > 0.5 M_{\odot}$ the Cowling approximation allows finding the oscillation frequency with an error in the range $\sim 15-30 \%$, which is not so bad in view of the enormous simplification in the calculations \footnote{We notice that the value of the parameter $\Delta$ is incorrectly labeled in Refs.  \cite{vasquez2010,vasquez2014,vasquez2014nach}. The true value employed in the calculations is obtained by multiplying the indicated value by a factor 3/2 (e.g. where it reads $\Delta = 75$ MeV it should read  $\Delta = 112.5$ MeV).}.

\section{Summary and conclusions}
\label{conclusions}

In this work we have performed a detailed study of the properties of the fundamental mode of pulsation of color flavor locked strange stars. Using an equation of state derived within the MIT bag model we have solved the general relativistic equations for non-radial pulsations and we have obtained the frequency and the gravitational damping time of the fundamental mode for all the allowed parametrizations of the equation of state. 

We find  that the  frequency of the fundamental mode is  in the range $\sim 1-3$ kHz. For a given choice of the EOS's parameters, $f$ varies by less than $\sim 30 \%$ in the mass interval from 0.5 $M_{\odot}$ to the maximum mass. The frequency depends on the EOS's parameters as follows (see Fig. \ref{frequency}): as $\Delta$ is increased keeping constant the other parameters the $f$ versus $M$ curves move downwards and to the right; as $B$ (or $m_s$) is increased keeping constant the other parameters the $f$ versus $M$ curves move upwards and to the left. 
The damping time is in the range $\sim 0.2-1$ s for CFL strange stars with $M > 0.5 M_{\odot}$. Keeping constant all other parameters the $\tau$ versus $M$ curves behave as follows (see Fig. \ref{damping}): if $\Delta$ is increased  the curves move upwards and to the right; if $B$ (or  $m_s$) is increased the curves move downwards and to the left.
Our results show that CFL strange stars can radiate in the optimal range for present gravitational wave detectors and that 
it is possible to constrain the EOS' parameters if the fundamental oscillation mode is observed and the stellar mass is determined.

We have also shown that the $f$-mode frequency can be fitted as a function of the square root of the average stellar density $\sqrt{M/R^3}$ by a single linear relation that fits quite accurately the results for all parametrizations of the EOS (see Eq. \eqref{fitting_f}). All results for the damping time  can also be fitted as a function of the compactness $M/R$  by a single  empirical relation given in Eq. \eqref{fitting_tau}. 
Therefore, if a given compact object is identified as a CFL strange star (e.g. from the photon emission properties of a bare surface) these two relations could be used to determine the mass and the radius from the knowledge of the frequency and damping time of the $f$ mode. 
{On the other hand, since the fittings for CFL strange stars are  different to typical hadronic fittings (see  Figs. \ref{fig:linear_f} and \ref{fig:linear_tau}) the observation of the $f$-mode could potentially help in the discrimination of CFL strange stars and hadronic stars. However, notice that the hadronic and CFL fittings tend to overlap for values above $M/R \sim$ 0.27 and above $\sqrt{M/R^3} \sim 0.06 \mathrm{km}^{-1}$ which is precisely the range of more massive stars, i.e. the more interesting from the observational point of view. This overlapping is in accordance with the results of Ref. \cite{vasquez2014}, where the fluid modes of hadronic, hybrid and strange stars were investigated within the Cowling approximation. According to  Ref. \cite{vasquez2014}, for masses above $\sim 1 M_{\odot}$ it is difficult to discriminate   hadronic, hybrid and strange stars based solely on the $f$-mode, because of the superposition of the curves. To discriminate them it is necessary to
perform a  combined comparison of  the $g$ mode  (for hybrid stars) and the first pressure  mode $p_1$. Nonetheless, we emphasize that if a given star is identified as a CFL strange star by another method (e.g. through its electromagnetic or neutrino emission) the $f$-mode can be used to determine the stellar mass and radius.  }

If excited, the fundamental mode is damped not only by gravitational radiation, but also by viscous mechanisms if the viscosity is large enough. 
In the case of unpaired quark matter the most effective damping reaction is the nonleptonic process $u + d \leftrightarrow u + s$. For a typical stellar oscillation frequency of $\sim 1$ kHz,  oscillations are viscously damped in fractions of a second due to the nonleptonic process \cite{wang1984,sawyer1989,madsen1993,madsen1992}, and can interfere in the determination of the gravitational damping time.   
However, for the CFL phase the picture changes drastically because the above mentioned nonleptonic reactions are strongly suppressed \cite{alford2007a}. The thermodynamic and hydrodynamic properties are rather determined by the massless superfluid phonons $\varphi$ and  thermally  excited  light  pseudo-Nambu-Goldstone bosons. At low temperatures, the contribution to bulk viscosity  from  phonons  alone is several orders of magnitude smaller than for unpaired quark matter \cite{manuel2007,escobedo2009,andersson2010}. 
However,  the  dominant  contribution to the bulk viscosity may come from  processes involving the neutral kaon $K^0$  \cite{alford2007a,alford2007b,andersson2010}. For  temperatures above a few MeV, the bulk viscosity $\zeta^{K^0}$  can become larger than for unpaired quark matter, but at  low temperatures  $\zeta^{K^0}$ is much less than that of unpaired quark matter (see also  \cite{Bierkandt2011}). As a consequence, for cold enough objects we may expect that the fundamental mode will be damped only by gravitational radiation, and the determination of the frequency and damping time of the fundamental mode could be a promising way to constrain the properties of CFL strange stars.

\section{Acknowledgements}
C. V\'asquez Flores acknowledges the financial support received from CAPES. G. Lugones
acknowledges the financial support received from FAPESP and CNPq.

\end{document}